\def\revtex{1}

\ifx\revtex\undefined

\documentclass[axioms,article,submit,oneauthor,pdftex]{Definitions/mdpi}

\firstpage{1}
\pubvolume{1}
\issuenum{1}
\articlenumber{0}
\pubyear{2021}
\copyrightyear{2021}
\datereceived{}
\dateaccepted{}
\datepublished{}
\hreflink{https://doi.org/} 

\Title{Interdimensionality}

\TitleCitation{Interdimensionality}

\Author{Karl Svozil $^{1}$\orcidA{}}

\AuthorNames{Karl Svozil}

\AuthorCitation{Svozil, K.}

\address[1]{%
$^{1}$ \quad Institute for Theoretical Physics, TU Wien, Wiedner Hauptstrasse 8-10/136, 1040 Vienna,  Austria; svozil@tuwien.ac.at; \url{http://tph.tuwien.ac.at/~svozil}}

\corres{Correspondence: svozil@tuwien.ac.at}

\abstract{In this speculative analysis, interdimensionality is introduced as the (co)existence of universes embedded into larger ones. These interdimensional universes may be isolated or intertwined, suggesting a variety of interdimensional intrinsic phenomena that can only be understood in terms of the outer, extrinsic reality.}



\usepackage{mathbbol} 
\begin{document}

\else
\documentclass[%
   reprint,
   twocolumn,
 showpacs,
 showkeys,
 preprintnumbers,
  nofootinbib,
 amsmath,amssymb,
 aps,
 pra,
  longbibliography,
 ]{revtex4-2}

\usepackage[dvipsnames]{xcolor}

\usepackage{mathptmx}

\usepackage{amssymb,amsthm,amsmath,bm}

\usepackage{tikz}
\usetikzlibrary{calc,decorations.pathreplacing,decorations.markings,positioning,shapes,snakes}

\usepackage[breaklinks=true,colorlinks=true,anchorcolor=blue,citecolor=blue,filecolor=blue,menucolor=blue,pagecolor=blue,urlcolor=blue,linkcolor=blue]{hyperref}
\usepackage{graphicx}
\usepackage{url}

\usepackage{iftex}
\ifxetex
%
%
\usepackage{fontspec}
\usepackage{fontspec}
\setmainfont{Garamond}
\setsansfont{Garamond}
\fi

\usepackage{mathbbol} 

\begin{document}

\title{Interdimensionality}

\author{Karl Svozil}
\email{svozil@tuwien.ac.at}
\homepage{http://tph.tuwien.ac.at/~svozil}

\affiliation{Institute for Theoretical Physics,
TU Wien,
Wiedner Hauptstrasse 8-10/136,
1040 Vienna,  Austria}

\date{\today}

\begin{abstract}
In this speculative analysis, interdimensionality is introduced as the (co)existence of universes embedded into larger ones. These interdimensional universes may be isolated or intertwined, suggesting a variety of interdimensional intrinsic phenomena that can only be understood in terms of the outer, extrinsic reality.
\end{abstract}


\maketitle

\fi

\section{A caveat: speculation and progress}

Rule inference is the process of hypothesizing a general rule or ``law'' from examples or ``phenomena''~\cite{go-67,angluin:83}.
The halting problem is the task to determine,
given an arbitrary computer program and an input,
whether the program will eventually halt or continue to run forever. It is generally provable unsolvable.
As the former rule inference problem can be reduced to the latter halting problem, it is generally provable unsolvable.
This constraint on induction has been coped with by the philosophy of science in a variety of ways:
Popper suggested that, instead of induction and verification, which appears to be a hopeless endeavor,
falsification might be a good demarcation criterion between science on the one hand,
and on the other hand ideology, sophisms, or, in a more frugal term,
bullshit~\cite{Frankfurt-OnBullshit}.
Lakatos responded by criticizing that, due to side assumptions and a vast `protective belt' of auxiliary hypotheses,
in many practical circumstances, falsification fails.
As a result, contemporaries seldom have a clue as to what might turn out to become a
progressive versus a degenerative research program~\cite{lakatos_1978}.
Kuhn observed that science may be characterized by brief iconoclastic periods of revolution,
followed by longer conformist periods of consolidation~\cite{kuhn}.
Feyerabend even challenged methodology as mythology and ideology akin to religious dogmas,
and suggested to keep science wide open and perform an ``exhaustive search'' of ideas by allowing
``anything'' to enter the scientific debate,
thereby imposing little methodologic restrictions~\cite{fey-philpapers2};
he also recommended a formal separation between state and science,
and lay judges for evaluation of success~\cite{feyerabend-defense} and the allocation of scientific funding.

In any case, there seems to be no convergence of conceptual progression.
Take gravity and celestial motion, for example:
the Ptolemaic system was expressed in terms of geometry. It was superseded by the Copernican revolution that later became based on Newtonian gravitational forces.
Later on, Newtonian gravity was replaced by the curved geometry of space-time of Einstein's theory of general relativity.
By analogy, it appears highly likely that our contemporaries would view any model superseding the present canon as utterly speculative,
if not outright nonsense.

Such historic perspective leads to greater liberty and openness of ideas, and yet this creativity needs to be guided and stimulated
by empirical findings and attempts to falsify consequences and claims. This amounts to an amalgam of the aforementioned ideas brought forward
in the philosophy of science, resulting in a sort of pragmatism that is well balanced between wild phantasy and empirical grounding.
Exactly how much of those ingredients are in order may greatly depend on the temperament and character of the individual researcher.

We, therefore, present the following considerations with a caveat to the reader,
as it trespasses far beyond any empirically verifiable physics of our time;
and yet at least some aspects of it might indicate or sketch the way to fruitful avenues of scientific modeling.
We hope that the following speculations are not too weird for the realistic, critical, and sober mind.
At best this could be seen as a vision of things to come.

\section{Definition}

Interdimensionality, or, by another naming, dimensional shadowing~\cite{sv4}---the ``emulation'' of a lowerdimensional configuration space
by a fractal subset of a higherdimensional
manifold---is the (co)existence and (co)habitation of parts or fragments of an ``outer'' space of ``higher'' extrinsic Hausdorff dimension~\cite{falconer2}
by some ``inner'' subspace entity that has a ``lower'' or equal intrinsic Hausdorff dimension.
One may imagine such a situation as a fractal of Hausdorff dimension $d$ embedded in a continuum,
such as the Hilbert space $\mathbb{R}^n$ or $\mathbb{C}^n$, with $d \le n$.
So, pointedly speaking, we might exist on a sort of Cantor set or
Menger sponge-like structure---fractals obtained by self-similar elimination of proper parts---of (almost) integer
Hausdorff dimension which is part of a high-dimensional super-verse.

Formally the Hausdorff dimension $d$ of a set $A \in \mathbb{R}^n$, defined via the $d$-dimensional Hausdorff measure,
is based on its ``umklapp'' property---the sudden change from measure value zero to infinity if the dimension parameter
is taken higher or lower than a unique value---as follows.
Suppose  $\cup_i F_i$ covers $A$, and suppose further that
there exists a limit in which all individual
constituents $F_i$ of this covering become infinitesimal
in diameter.
Then the Hausdorff measure $\mu_d$, and a unique dimensional parameter $d$ called the
Hausdorff dimension
is
\begin{equation}
 \begin{aligned}
\mu_\delta (A)=
 &
\lim_{\epsilon \rightarrow 0+} \inf_{\{ F_i\} }
\left\{ \sum_i\big(\text{diam }F_i\big)^\delta \middle|
 \right. &\\
 &
\; \delta \in \mathbb{R},
\; \delta > 0,
\;
\cup_iF_i\supset A,
\; \big(\text{diam }F_i\big) \le \epsilon
 \Big\},&
 \end{aligned}
\end{equation}
where the infimum is over all countable $\epsilon$-covers $\{ F_i\}$ of $A$;
with the dimension $d$ as an ``umklapp'' parameter of
\begin{equation}
\mu_\delta (A)=
\begin{cases}
0 & \text{ if } \delta > d, \\
\infty &  \text{ if } \delta < d.
\end{cases}
\end{equation}
That is, the Hausdorff dimension $d$ is the unique dimensional parameter at which the measure
$\mu_\delta$ as a function of the dimensional parameter value $\delta$ smaller or larger than $d$ is infinite or vanishes, respectively.
Note that the diameter ``$\text{diam}$'' presupposes the notion of a distance defined via a metric.
For self-similar fractal sets, the
capacity dimension $c$ is defined by
\begin{equation}
c=\lim_{\epsilon \rightarrow 0+}\;\log \left[ n(\epsilon )\right] /\log \left(
\epsilon^{-1}\right),\end{equation}
where $n(\epsilon )$ is the number of segments
of length $\epsilon$,
equals the Hausdorff dimension $d$.

An example of a set of integer dimension $m$ embedded into an outer space $\mathbb{R}^n$  with $n > m$
is the set  whose (contravariant) coordinates with respect to some (covariant) basis $\mathbb{R}^n$ is given by
\begin{equation}
 \begin{aligned}
\left\{
\Big(
 \right.
 &
x_1,x_2,\ldots ,x_m,
 \\
 &
 \; \left.
r_1(x_1,x_2,\ldots ,x_m),  \ldots , r_{n-m}(x_1,x_2,\ldots ,x_m)
\Big)^\intercal \middle|
 \right. \\
 &
 \; \;
 x_i, r_j(x_1,x_2,\ldots ,x_m) \in \mathbb{R}
 \Big\}
,
\end{aligned}
\end{equation}
where $r_i(x_1,x_2,\ldots ,x_m)$, $1 \le i \le n-m$ are some total, possibly constant or random, choice functions.

For most practical operational purposes~\cite{sv1,sv2}  the intrinsic perception of the dimensionality of such shadowed, interdimensional object
might effectively remain that of a ``solid continuum'' of that intrinsic (Hausdorff) dimension.
It may not be too unreasonable to compare this to the common notion of  ``emptiness of space in-between point particles''
constituting solid physical objects,
or the ``perceived continuous motion'' from individual still frames~\cite{Wertheimer-12,goldstein-Brockmole}.

There are some findings  consistent such speculations:
For instance,  associated with every integer-dimensional
regular rectifiable $m$-dimensional fractal
embedded in ${\Bbb R}^n$ there exists a locally defined tangential $m$--dimensional
vector subspace of ${\Bbb R}^n$~\cite{federer1,falconer2}.
Even for non-integer-dimensional fractals, integer-dimensional tangent spaces may be ``good'' approximations for all practical physical purposes.

Further examples for cohabitation of continua that need not involve fractals are paradoxical decompositions, such as
Vitali's partition of the unit interval and the decomposition of the sphere by Hausdorff~\cite{wagon1}.
If we relax the definition of dimension we may also speak of (dense) ``scattered'' point sets ``inhabiting'' the continuum.
The variations may be manyfold; for instance, one may consider partitions or intertwined subsets of continua.
And one may not even deal with extrinsic continua but with general sets that allow some form of intrinsic embeddings.

\begin{figure}
\begin{center}
\begin{tabular}{cc}
\includegraphics[width=0.2\textwidth]{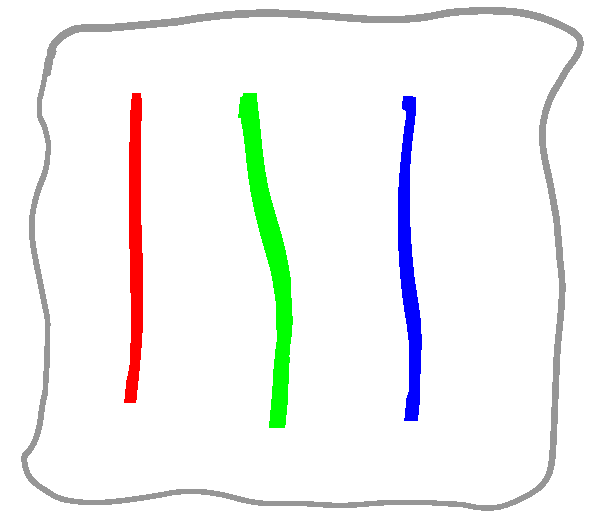}
\quad & \quad
\includegraphics[width=0.2\textwidth]{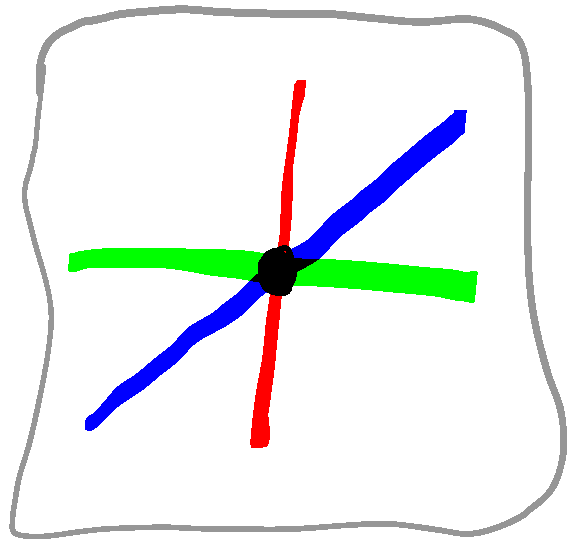}
\\
(a)&(b)\\
\end{tabular}
\end{center}
\caption{\label{2021-inter-f1}
Schematic drawing of interdimensional configurations that are (a)~isolated or (b)~intertwine, as seen from some outer, embedding space.
}
\end{figure}

Let us finally review two almost trivial examples of an arbitrary number of one-dimensional subspaces of $\mathbb{R}^2$,
as schematically depicted in Figure~\ref{2021-inter-f1}.
The first one is a collection of parallel lines.
The second one is a star-shaped configuration intertwining in the origin, spanned by respective mutually distinct unit vectors.
In the latter case, the only way of ``flatlanders''~\cite{abbott-flatland} living on different subspaces to communicate with each other is through
a single point---the origin.

In general, fractals need not be regular and rectifiable and of integer dimension.
Rather they may be ``cloud-like shapes'', with ``scattered'' holes and gaps. Those gaps will not be perceived intrinsically.
Indeed one may speculate that this situation gives rise to a metric that essentially mimics curvature~\cite{svozil-2017-fg}.

Fractal theory has inspired and evolved into many innovative, useful and interesting applications, especially in new materials and nanostructures.
Such important developments can lead us to new views of, and physical means related to, dimensionality~\cite{Hill-2017,Mitic-2019}.

As the aim is the provision of a very general analysis that is unconstrained by the technicalities of specific models,
no concrete theory is discussed.
Nevertheless, it might be not too far-fetched to just briefly mention some potential connections between interdimensionality
and various paradigms in modern particle physics and cosmology.
Some of these involve the description of a volume of space as conceptualized by
holographic principles, such as the AdS/CFT correspondence related to D-branes in string theory,
or the ekpyrotic models relying on string theory, branes and extra ``hidden'' dimensions.
Other scenarios in the context of the theory of general relativity involve traversable wormholes
(aka Einstein-Rosen bridges) linking disparate points in spacetime.

\section{Disjoint and intertwining shadows}

To proceed to interdimensional motion we need to consider intertwining areas of interdimensionality.
The simplest nontrivial case is the one schematically depicted in Figure~\ref{2021-inter-f1}(b) in which all universes share a single point of communication.
Of greater interest might be a situation in which an entire region of space is shared.
One might think also of a ``small'' fraction of a universe ``traversing'' another universe; such that,
compared to the overall extension of these universes this common share appears like the tip of an iceberg.

\section{Interdimensional motion}

Interdimensional motion is the motion of some ``inner'' intrinsic subspace in the ``outer'', extrinsic space.
If two inner spaces are involved it may happen that certain limits of motion,
such as continuity or maximal speed, that are valid in one subspace, can be breached and overcome by another subspace.
In what follows some scenarios will be discussed.
We shall adopt the following notation: inner ``intrinsic'' subspaces will be denoted by $\textsf{\textbf{M}}$ and $\textsf{\textbf{N}}$.

\begin{figure}
\begin{center}
\begin{tabular}{ccc}
\includegraphics[width=0.13\textwidth]{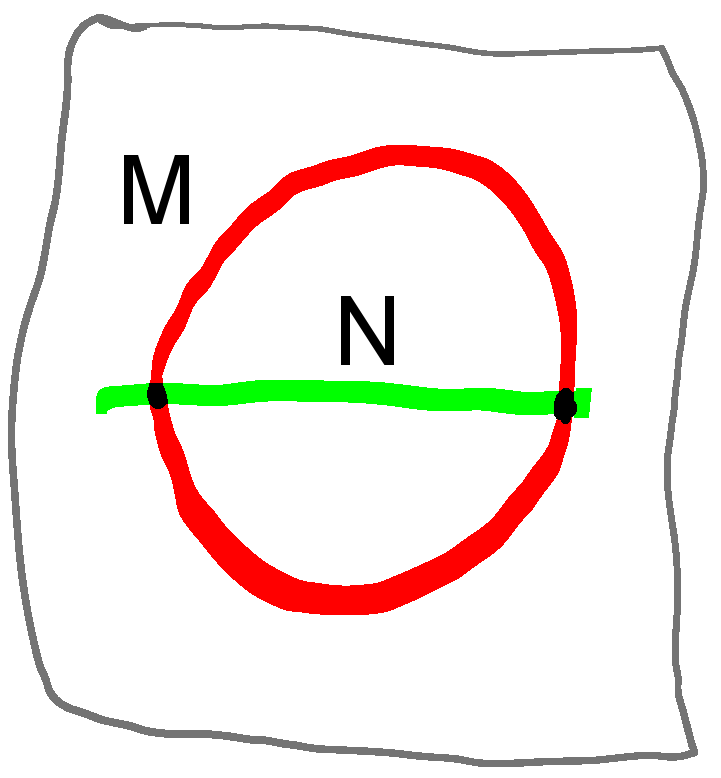}
\quad & \quad
\includegraphics[width=0.13\textwidth]{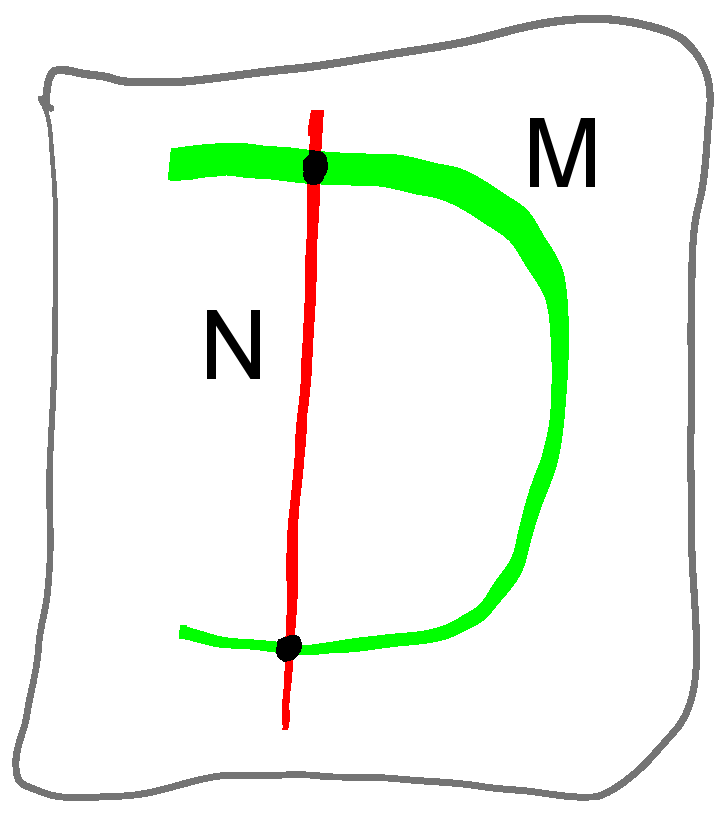}
\quad & \quad
\includegraphics[width=0.13\textwidth]{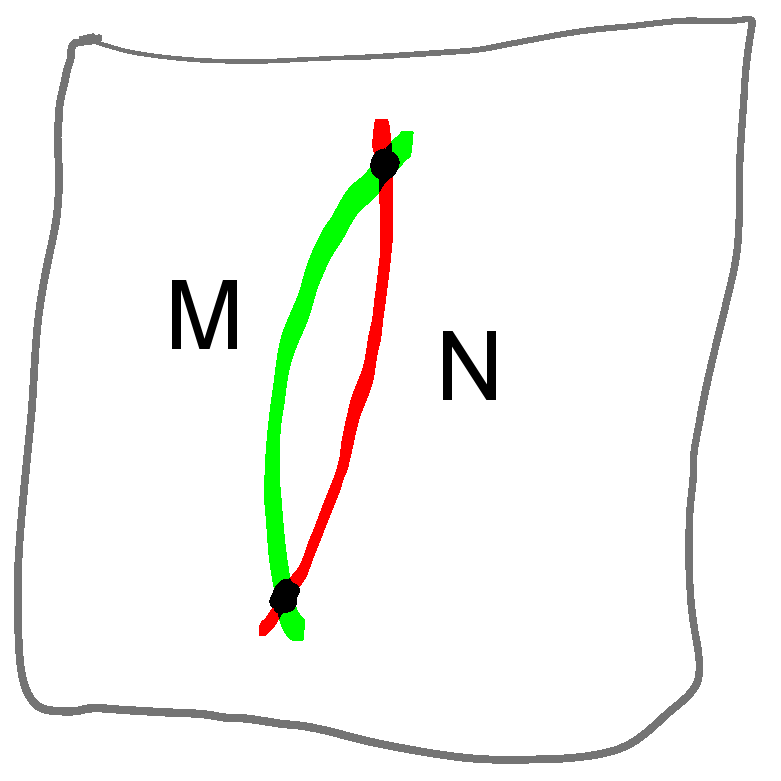}
\\
(a)&(b)&(c)\\
\end{tabular}
\end{center}
\caption{\label{2021-inter-f2}
Schematic drawing of worldlines of interdimensional motion, as seen from the outer, embedding space: (a) periodic, (b) shortcut, and (c) coevolution.
}
\end{figure}

Let us discuss this by considering a simple example of a rotating point, as schematically drawn in Figure~\ref{2021-inter-f2}(a).
From the point of view of  $\textsf{\textbf{M}}$ the rotation in $\textsf{\textbf{N}}$ is observed as periodic (dis)appearances of some object rotating in $\textsf{\textbf{M}}$.

Another ``wormhole''-like scenario schematically drawn in Figure~\ref{2021-inter-f2}(b) is a ``bend'' or ``curved'' (relative to the exterior ``outer'' continuum)  reference frame $\textsf{\textbf{M}}$ that is
intermittantly accessed from $\textsf{\textbf{N}}$. Suppose that the propagation speed limit for motion is the same $c_\textsf{\textbf{M}}=c_\textsf{\textbf{N}}$ in both frames.
Then the object appears to be traveling with a velocity greater than this limit velocity in $\textsf{\textbf{M}}$ because of the ``shortcut'' access through $\textsf{\textbf{N}}$.

Still another scenario schematically drawn in Figure~\ref{2021-inter-f2}(c) is one in which $\textsf{\textbf{N}}$ allows for faster that $\textsf{\textbf{M}}$--light motion---that
is, $c_\textsf{\textbf{M}} \ll c_\textsf{\textbf{N}}$---and this property is used to access regions in $\textsf{\textbf{M}}$ through motion in $\textsf{\textbf{N}}$ that
appear spece-like separated in $\textsf{\textbf{M}}$'s frame of reference.

\subsection{Interdimensional chronology protection}

In these and similar situations no issues with respect inconsistent evolution, in particular, time paradoxes, arise.
Because whatever relative space-time reference frames are operationally constructed~\cite{svozil-relrel}
in $\textsf{\textbf{M}}$ and $\textsf{\textbf{N}}$,
the ``outer'' extrinsic space in which both $\textsf{\textbf{M}}$ and $\textsf{\textbf{N}}$ are embedded regulates the phenomenology.

Indeed, from an extrinsic, ``God's eye view'' of the outer space there is no consistency issue
because the evolution seen from this ``global'' comprehensive perspective never yields or allows inconsistent phenomena.
Concerns raised by intrinsic space-time frames generated with the means available
in $\textsf{\textbf{M}}$ and $\textsf{\textbf{N}}$ are merely epistemic, and means relative to the devices and conventions
(such a for synchronizing clocks)
available to the inhabitants of  $\textsf{\textbf{M}}$ and $\textsf{\textbf{N}}$.

This results in an interdimensional scheme of chronology protection based on the epistemic relativity of reference frames.
At the same time, from an ``outer'' (ontological if you accept the term) point of view those frames are
``bundled together'' through the coembedding and cohabitation of some outer space.

There are similarities between the consistency of observable phenomena regarding the higher-dimensional bulk space
and the consistent histories approach to Many Worlds models~\cite{Carr-2007}.
Both involve multiple ``merging'' paths.

\subsection{Examples of dimensional relativity}

The following examples closely follow the scenarios schematically depicted in Figures~\ref{2021-inter-f2}(b,c).
They have some similarities to ballistic missiles
that avoid limitations of velocity from atmospheric drag (friction) by leaving and re-entering Earth's atmosphere,
or are analogs of supercavitation---the formation of vapour bubbles in a liquid caused by flow around an object, allowing minimal friction movement inside liquids at nearly sound speeds.

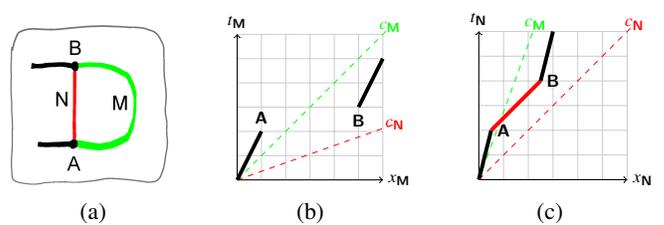
\begin{figure}
\begin{center}
\begin{tabular}{ccc}
\includegraphics[width=0.13\textwidth]{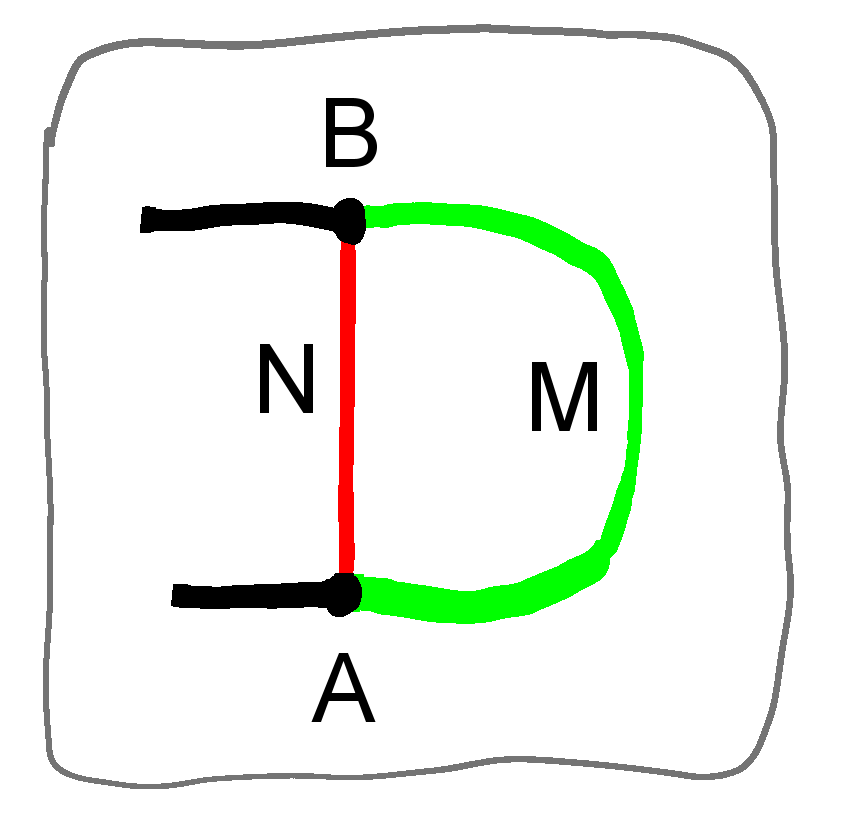}
\quad & \quad
\resizebox{.15\textwidth}{!}{
\begin{tikzpicture}  [scale=0.45]
  \draw[thin,gray!40] (0,0) grid (6,6);
  \draw[->] (0,0)--(6,0) node[right]{$x_{\textsf{\textbf{M}}}$};
  \draw[->] (0,0)--(0,6) node[above]{$t_{\textsf{\textbf{M}}}$};

  \draw[line width=1pt,dashed,thin,green](0,0)--(6.3,6.3) node {$c_{\textsf{\textbf{M}}}$};
  \draw[line width=1pt,dashed,thin,red](0,0)--(6.5,2.3) node {$c_{\textsf{\textbf{N}}}$};

  \draw[line width=2pt,black](0,0) -- (1,2) node[above]{${\textsf{\textbf{A}}}$};
  \draw[line width=2pt,black](5,3) node[below]{${\textsf{\textbf{B}}}$} -- (6,5);
\end{tikzpicture}
}
\quad & \quad
\resizebox{.15\textwidth}{!}{
\begin{tikzpicture}  [scale=0.45]
  \draw[thin,gray!40] (0,0) grid (6,6);
  \draw[->] (0,0)--(6,0) node[right]{$x_{\textsf{\textbf{N}}}$};
  \draw[->] (0,0)--(0,6) node[above]{$t_{\textsf{\textbf{N}}}$};

  \draw[line width=1pt,dashed,thin,green](0,0)--(2.3,6.3) node {$c_{\textsf{\textbf{M}}}$};
  \draw[line width=1pt,dashed,thin,red](0,0)--(6.3,6.3) node {$c_{\textsf{\textbf{N}}}$};

  \draw[line width=2pt,black] (0,0) -- (0.5,2);
  \node at (1,2){${\textsf{\textbf{A}}}$};
  \draw[line width=2pt,red] (0.5,2) -- (2.5,4);
  \draw[line width=2pt,black] (2.5,4)  -- (3,6);
  \node at (3,4) {${\textsf{\textbf{B}}}$};
\end{tikzpicture}
}
\\
(a)&(b)&(c)\\
\end{tabular}
\end{center}
\caption{\label{2021-inter-f3}
Schematic drawing of (a) worldlines of interdimensional ``jump'' motion, as seen from the outer, embedding space:
(a) ``dive'' into $\textsf{\textbf{N}}$ at $\textsf{\textbf{A}}$, reappearance at $\textsf{\textbf{B}}$;
(b) space-time diagram as seen from intrinsic coordinates in $\textsf{\textbf{M}}$;
(c) space-time diagram as seen from intrinsic coordinates in $\textsf{\textbf{N}}$.}
\end{figure}

The first example, depicted in Figure~\ref{2021-inter-f3},
shows an interdimensional dive into a dimension that allows higher velocities, or rather traversals of space per time,
in $\textsf{\textbf{M}}$ through ``jump'' into another dimension $\textsf{\textbf{N}}$,
thereby creating a shortcut from two space-time points $\textsf{\textbf{A}}$ to $\textsf{\textbf{B}}$.
This is different from breaking the intradimensional warp barrier
by hyper-fast solitons in Einstein-Maxwell-plasma theory~\cite{Lentz2021}
as it employs dimensional capacities that are not bound by intradimensional motion.

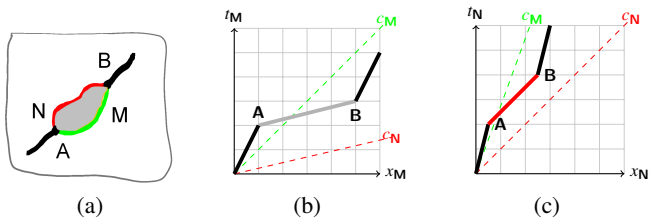
\begin{figure}
\begin{center}
\begin{tabular}{ccc}
\includegraphics[width=0.13\textwidth]{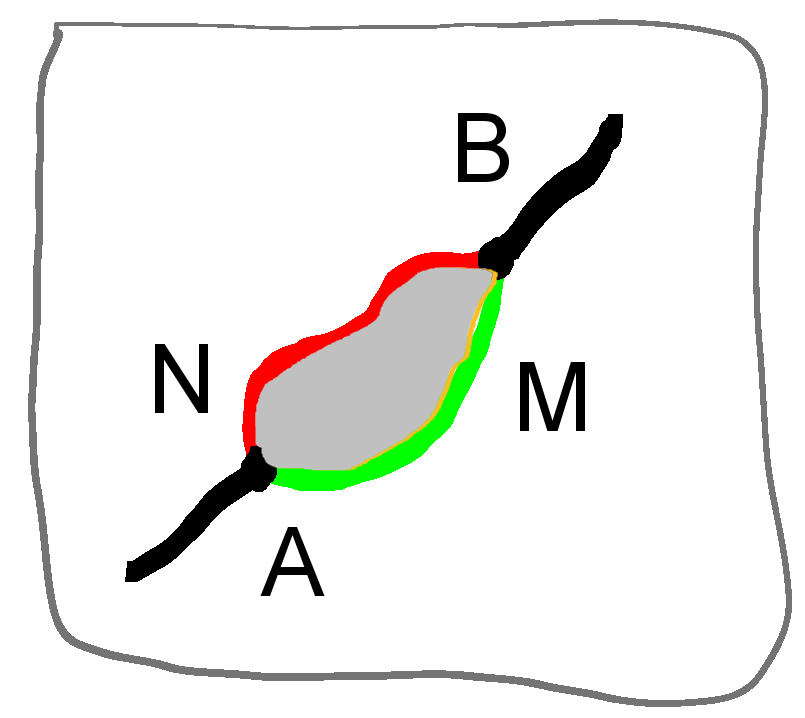}
\quad & \quad
\resizebox{.15\textwidth}{!}{
\begin{tikzpicture}  [scale=0.45]
  \draw[thin,gray!40] (0,0) grid (6,6);
  \draw[->] (0,0)--(6,0) node[right]{$x_{\textsf{\textbf{M}}}$};
  \draw[->] (0,0)--(0,6) node[above]{$t_{\textsf{\textbf{M}}}$};

  \draw[line width=1pt,dashed,thin,green](0,0)--(6.3,6.3) node {$c_{\textsf{\textbf{M}}}$};
  \draw[line width=1pt,dashed,thin,red](0,0)--(6.5,1.51) node {$c_{\textsf{\textbf{N}}}$};

  \draw[line width=2pt,gray!60] (1,2) -- (5,3);
  \draw[line width=2pt,black](0,0) -- (1,2) node[above]{${\textsf{\textbf{A}}}$};
  \draw[line width=2pt,black](5,3) node[below]{${\textsf{\textbf{B}}}$} -- (6,5);
\end{tikzpicture}
}
\quad & \quad
\resizebox{.15\textwidth}{!}{
\begin{tikzpicture}  [scale=0.45]
  \draw[thin,gray!40] (0,0) grid (6,6);
  \draw[->] (0,0)--(6,0) node[right]{$x_{\textsf{\textbf{N}}}$};
  \draw[->] (0,0)--(0,6) node[above]{$t_{\textsf{\textbf{N}}}$};

  \draw[line width=1pt,dashed,thin,green](0,0)--(2.3,6.3) node {$c_{\textsf{\textbf{M}}}$};
  \draw[line width=1pt,dashed,thin,red](0,0)--(6.3,6.3) node {$c_{\textsf{\textbf{N}}}$};

  \draw[line width=2pt,black] (0,0) -- (0.5,2);
  \node at (1,2){${\textsf{\textbf{A}}}$};
  \draw[line width=2pt,red] (0.5,2) -- (2.5,4);
  \draw[line width=2pt,black] (2.5,4)  -- (3,6);
  \node at (3,4) {${\textsf{\textbf{B}}}$};
\end{tikzpicture}
}
\\
(a)&(b)&(c)\\
\end{tabular}
\end{center}
\caption{\label{2021-inter-f4}
Schematic drawing of (a) worldlines of interdimensional forced, continuous motion, as seen from the outer, embedding space:
(a) until $\textsf{\textbf{A}}$ and from $\textsf{\textbf{B}}$, the motion is dominated by contraints on the
velocity $v_\textsf{\textbf{N}}$, and inbetween $\textsf{\textbf{A}}$ and  $\textsf{\textbf{B}}$ the velocity $c_\textsf{\textbf{N}}$ dominates;
(b) space-time diagram as seen from intrinsic coordinates in $\textsf{\textbf{M}}$;
(c) space-time diagram as seen from intrinsic coordinates in $\textsf{\textbf{N}}$.}
\end{figure}

The second example, depicted in Figure~\ref{2021-inter-f4},
shows an interdimensional ``drag'' motion that uses a dimensional motion in $\textsf{\textbf{N}}$
whose velocity exceeds that of the normal signal velocity in $\textsf{\textbf{M}}$.
As already mentioned in both of these cases consistency is guaranteed by the overall consistency in the outer embedding space.

\section{Further speculations}

Let us conclude this speculative article with some speculative thoughts.
The first one is on limits to isolating the dimensions from one another, from ``keeping them apart''; in particular,
in the event of some catastrophic occurrence.
It may well be that the domain of dimensional intersections may increase,
as such events may dominate and spread to larger parts of the ``outer'' space.

Secondly, interdimensionality can be compared to computer simulations,
with interfaces between such universes serving as intertwining regions.
The difference between virtual reality (exchanges) and (intertwining) interdimensionality is the
emphasis on measure-theoretic aspects in the latter case.

Let me again point out that the matters discussed here must be considered highly speculative,
and far from a fully developed formal theory.
Nevertheless, it is our conviction that, to progress,
science has to expand and explore  a great variety of options,
even if they appear remote to the contemporary mind.

\ifx\revtex\undefined

\funding{This research was funded in whole, or in part, by the Austrian Science Fund (FWF), Project No. I 4579-N. For the purpose of open access, the author has applied a CC BY public copyright licence to any Author Accepted Manuscript version arising from this submission.}


\conflictsofinterest{The author declares no conflict of interest.
The funders had no role in the design of the study; in the collection, analyses, or interpretation of data; in the writing of the manuscript, or in the decision to publish the~results.}

\else

\begin{acknowledgments}

This research was funded in whole, or in part, by the Austrian Science Fund (FWF), Project No. I 4579-N. For the purpose of open access, the author has applied a CC BY public copyright licence to any Author Accepted Manuscript version arising from this submission.

The author declares no conflict of interest.
\end{acknowledgments}

\fi

\ifx\revtex\undefined

\end{paracol}
\reftitle{References}



\else


%

\fi
\end{document}